\documentclass[12pt]{article}
\usepackage{amsmath}
\usepackage{tensor}
\newcommand{\be}{\begin{equation}}
\newcommand{\ee}{\end{equation}}
\newcommand{\bea}{\begin{eqnarray}}
\newcommand{\eea}{\end{eqnarray}}
\newcommand{\nn}{\nonumber}

\newcommand{\w}{\mbox{$\omega$}}
\newcommand{\g}{\mbox{$\gamma$}}

\newcommand{\p}{\mbox{$\partial$}}

\newcommand{\vv}{\vec{v}}
\newcommand{\va}{\vec{a}}

\begin{document}

\section*{Radiation Reaction Force on a Charged Particle.}

{\bf H. Fearn$^1$ and J. Bengtsson$^2$}\\

\noindent
{\small $^1$ Physics Department, California State University Fullerton\\
800 N. State College Blvd, Fullerton CA 92834.\\
{\bf email:} hfearn@fullerton.edu\\

\noindent
{\small $^2$ Brookhaven National Laboratory\\
Upton, NY 11973-5000\\
{\bf email:} bengtsson@bnl.gov \\

\noindent
{\bf Keywords:} Radiation Reaction, Poynting--Robertson force, Lorentz--Abraham--Dirac equation\\
\noindent
{\bf PACS:} 03.50.De  04.04.N1  41.60.-m\\

\vspace{ 0.25in}

\noindent {\large {\bf Abstract}}: This paper brings together
several works describing the force acting on a dust particle, an
atom or charge due to the radiation emitted by it, or surrounding
it. It is shown that you can easily generalize the power radiated
from the classical expression by Larmor to the general
relativistic expression by Li\'enard. The same method could be
used to generalize the force expression (from classical to
relativistic) if a well known term is added to the standard
Abraham-Lorentz force term with $ \dot{a}$. This addition is the
Poynting--Robertson term, seen mostly in astrophysics and usually
missing from texts in electromagnetism. With this term added, it
takes into account the rate of change of mass-energy $dm/dt\sim
c^{-2}d \varepsilon/dt$ of the particle and makes the
generalization to the relativistic force formula straightforward.\\

\subsubsection*{Introduction}


Let us begin with a brief discussion of when a particle radiates and why. Then establish whether a radiating particle
will experience a radiation reaction force.\\

\noindent
A charged particle will radiate when it is accelerated. It may or may not
experience a drag force (radiation reaction) from this. For a non-relativistic particle,
the usual radiation power and reaction force is given by Larmor \cite{larmor} and Abraham \cite{abraham} to be,
\be
P_{\mbox{\tiny Rad}}= \frac{ 2}{3} \frac{ q^2} {c^3} a^2 \;\;\;, \mbox{           } \mbox{           }
F_{\mbox{\tiny Rad}} = \frac{ 2}{3} \frac{ q^2} {c^3} \dot{a} \nn
\ee
where $q $ is the charge, $\dot{a}$ is the rate of change of acceleration and $c$ is the velocity of light in a vacuum.
It appears that when a charge has constant acceleration it radiates power but has no radiation reaction force.
As in the case where a charge particle
moves in a hyperbolic motion, the particle undergoes constant acceleration and radiates continuously and yet
experiences no drag force. This was explained by Fulton and Rohrlich in 1960 \cite{fulton} in terms of
conservation of energy.
It is necessary to take into account the mechanical energy of the particle, the radiated energy and the ``self-energy",
which is the energy in the field surrounding the particle. This self energy does not need to be only velocity dependent
and can be different on the incoming and outgoing legs of the hyperbolic motion.
The difference in energy between in the incoming and
outgoing legs of the journey is enough to account for the radiation emitted. In order to see the radiation it
was necessary to compute the power radiated from one part of the motion taken at a distance $ct$, calculate
power emitted through a sphere of that radius, then allow
$t \rightarrow \infty$. So, to see the radiated energy in a uniform, non-periodic situation
boundary conditions at infinity become important.\\

\noindent
The physics implies that one would expect radiation and a radiation reaction force in all non-uniform acceleration
regardless of whether the motion is periodic or not. Furthermore, one would expect
radiation to be emitted in all uniform acceleration conditions but whether the particle experiences a radiation
reaction drag force is only a matter of the boundary conditions. For periodic motion there will be a drag
(this is how the radiation reaction force term was initially derived) for
non-periodic motion there may not be a drag force, eg. hyperbolic motion.
The radiation reaction of a \emph{charged particle } can be thought of as the particle interacting with its own
radiation thus a periodic motion of velocity $v<c$ is helpful to account for the particles interaction with
light emitted at c. (See for example Griffiths \cite{griffiths}.) The radiation should also include the ``self-energy"
or near field (sometimes called the velocity or induction field) that stays near the particle and does not
radiate out to infinity.\\

\noindent
A charged particle can radiate if it moves faster than the speed of light in a material (other than vacuum),
this is the well known \emph{Cerenkov radiation}. A charged particle will also radiate if the particle moves across a
boundary from one material to another, at the intersection there will be \emph{transition radiation}. Both
effects are well known and can be found in standard text books. In the latter two cases the particle need not
be accelerating they can be moving at a constant speed. These particles are thought not to experience a drag force due to
the radiation. Usually these particles will slow down due to collisions with other atoms but then the Bremsstrahlung
causes the slowing and scattering, where deceleration or change of acceleration occurs.\\

\noindent
A non charged particle (dust) can absorb heat and re-emit
in the process of cooling. This is in fact how the Poynting--Robertson drag term was discovered and will be
addressed in the next section.

\subsubsection*{The Poynting--Robertson Drag Force.}

In 1903 J.\ H.\ Poynting \cite{poy} published work in the
Philosophical Transactions of the Royal Society entitled
``Radiation in the solar system. Its effects on temperature and
its pressure on small bodies". One of the more interesting results
of the paper was the radial and tangential force on the dust
particle due to radiation pressure from the sun and reradiation of
heat(cooling) from the particle, as seen from the suns rest frame.
The drag force derived by Poynting was of the form, \be
\label{eq_poynting} F_{\mbox{\tiny Drag}} = -\frac{v}{c^2} R = -
\frac{v}{c^2} \frac{d \varepsilon}{dt} = -v \frac{dm}{dt} \ee
where $c$ is the velocity of light, $R$ is the power of the
radiation, $\varepsilon$ is the energy, $m$ is the mass and $v$ is
the velocity of the particle. Poynting \cite{poy} actually derived
a third of this term for both the radial and the tangential
components. The radial force, due to radiation pressure, pushes
the particle away from the sun. The tangential component gives
rise to a decrease in angular momentum and hence the dust particle
would spiral into the sun. Whether the particle spirals inwards or
outwards depends on the size of the particle and initial
conditions.

\noindent
In 1912 Larmor \cite{larmor}, addressed the International Congress of Mathematicians in Cambridge
and showed that Poynting had forgotten to take into account the Doppler factor, for motion in
the rest frame of the sun. Larmor stated that the force required to change the momentum is of
the form,
\be
\frac{d}{dt} (m \vv ) = m \frac{ d \vv}{dt} + \frac{ dm}{dt} \vv \nn . \\
\ee The latter term involving the velocity as a factor is
$\frac{d}{dt} (\varepsilon/c^2) v = -R v/c^2 $. This agrees with
the derivation using the Doppler factor $(1-v/c)$. Larmor was also
concerned at the time that this cooling effect could act as a
brake on a particle in the depths of space, thereby allowing
motion to be detected in contradiction with the theory of special
relativity and thus the laws of electrodynamics.  Page \cite{page}
later showed that a particle does not suffer a retardation caused
by its own radiation, a fact that was later admitted by Larmor in
a postscript to Poyntings Collected papers \cite{poy} p 757 where
he states,

\begin{quote}
``...the remarkable result seems to be established that an isolated body
cooling in the depths of space would not change its velocity through the
aether, the retardation due to the back thrust of radiation issuing from it
being just compensated by increase of velocity due to momentum conserved with
diminishing mass".
\end{quote}

\noindent
Larmor continues,
\begin{quote}
``But for Poynting's particle describing a planetary orbit the radiation
from the Sun comes in, which restores the energy lost by radiation from the
particle, and so establishes again the retarding force [$-Rv/c^2$ ]".
\end{quote}

\noindent
All of the above discussion was summarized by Robertson \cite{rob} in 1937. The drag force was still not widely known
until Roberston drew attention to it and explained the effect. Finally the drag became known as the Poynting--Robertson
force, and this can be found in modern texts \cite{noon}. Burns, Lamy and Soter \cite{burns}, present
an up-to-date derivation of the Poynting--Robertson force and generalize to a non perfectly absorbing particle.
To conclude this section we summarize their ``perfect absorber" results here.
The drag force due to solar radiation can be addressed as two distinct terms;\\

\noindent
(a) A radiation pressure term; due to the momentum in the suns radiation hitting the particle and
pushing it away from the sun.\\
\noindent
(b) A mass-loading drag; due to the effective mass-energy loss from the particle as it re-radiates the incident
energy.\\

\noindent
Let the suns energy flux be given by the Poynting vector $S$. Let the dust particle have a cross-sectional area $A$
and be perfectly absorbing. Then the energy absorbed by the dust particle per unit time
 is $SA$. If the particle is moving at speed $v$
relative to the sun then we must replace $S$ by $S' = S(1 - \dot{r}/c)$ where $\dot{r} = \vv \cdot \hat{S}$ is the radial
velocity of the particle away from the sun. The momentum removed per second from the suns rays, by the dust particle,
is $S' A/c$ which is the radiation pressure force. The absorbed energy flux is re-radiated by the particle. In the rest
frame of the particle, the radiation emitted is almost isotropic so there is no net force on the particle in its own rest frame.
However, the re-radiation is equivalent to an energy loss rate $S'A/c^2$, which has velocity $\vv$ when viewed from the rest frame
of the sun. In the solar rest frame, the particle has a drag force of $ - S'A \vv/c^2 $. Since the dust particle is losing
momentum, while its mass is conserved, the particle is decelerated. The momentum loss per unit time is,
\bea
m \dot{v} &=& \frac{S' A}{c} \hat{S} - \frac{S' A}{c^2} \vv \nn \\
 &=& \frac{SA}{c}\left[ \left( 1 - \frac{ \dot{r} }{c} \right)\hat{S}  - \frac{\vv}{c} \right]  .
\eea
The last term on the right is the Poynting--Robertson drag and equivalent to the Robertson 1937 result.
This definition clearly allows for a dust particle \cite{poy}, but can also be used for
a dipole oscillator \cite{einhopf}, a 2--level atom, \cite{einstein}  or an
electron \cite{abraham}, \cite{laue} and \cite{dirac}. \\

\noindent
Historically, this form of radiation reaction was first developed by Abraham 1903 \cite{abraham} for an
electron. Similar force terms arise in the relativistic LAD (Lorentz, Abraham and Dirac) equation
see von Laue (1909) \cite{laue}, Dirac \cite{dirac}, Pauli's book \cite{pauli} and
Lorentz \cite{lorentz}, who quotes Abraham's 1903 result on page 31 of his Dover book. \\

\noindent
In the case of the point charge, the power radiated $R$ takes the usual Larmor form
\be
R = \int S_{\mbox{\tiny Rad}} \cdot d\sigma = -\frac{2}{3} \frac{e^2}{c^3}a^2
\ee
where $\sigma$ is an enclosed spherical area, $a$ is the
acceleration of the particle and $S_{\mbox{\tiny Rad}}$ is the time average
Poynting vector of the radiation emitted from the particle.\\

\noindent
Typically, the radiation reaction force is accredited
to Abraham and Lorentz and given by
\be
F_{\mbox{\tiny Rad}} = \frac{2}{3} \frac{e^2}{c^3} \dot{a}
\ee
where $a$ is the acceleration of the
particle (usually electron or point charge). See Jackson \cite{jack}. It
should be noted that this force varies as $1/c^3$ whereas the
Poynting--Robertson force varies as $1/c^5$. The Poynting--Robertson
force might be considered a small correction to the usual
radiative reaction, but without it, the generalization to
relativistic expression is not obvious.

\noindent
The main result of this paper is a generalization of the regular radiation reaction force
by adding a small term $\delta F$ to $F_{Rad}$. The $\delta F$ is found
equal to the Poynting--Robertson drag force. The new classical radiation reaction would become
\be
\vec{F}_{\mbox{\tiny Rad}} = \frac{2 e^2}{3 c^3} \left[ \dot{\va} + \frac{\vv}{c^2} a^2 \right]
\ee
We also show how this result can be simply derived in both the classical and relativistic cases.
In the following section we derive the result and generalize to the relativistic case. \\

\noindent
In subsequent sections we
give multiple examples of how the Poynting--Robertson term is the only radiative drag term available to account
for deceleration in a variety of physical situations when the acceleration is constant $\dot{a} = 0$.

\subsubsection*{Generalization of the Radiation Reaction Force}

For a lucid interpretation of the Lorentz-Dirac equation for a charged object, see A. D. Yaghjian's
analysis of the ``Relativistic Dynamics of a charged sphere", \cite{yag}.
In this section we would like to show how one can easily generalize power
radiated by a point charge to the fully relativistic form. We then generalize the
classical formula for radiation reaction force with the addition of the Poynting--Robertson term.\\

\noindent
Finally it is shown how easily this force expression can be generalized to the fully relativistic form.\\

\noindent
In teaching electromagnetism for many years you see tricks in
generalizing the non-relativistic power formula to the
fully covariant relativistic formula via changing
 $a^2 \rightarrow a^{\alpha} a_{\alpha}$ where $\vec{a}$ is the classical
acceleration 3--vector and $a^{\mu} = du^{\mu}/d\tau $ is the
4--acceleration. It is quite easy to show using
$x^{\mu}=(ct,\vec{x})$, metric ${+,-,-,-}$, the time dilation equation
$dt = \gamma d\tau$ and $u^{\mu} = (\gamma c, \gamma \vec{v})$ that
\be
a^{\mu} = \left( \gamma^4 \frac{\vec{v} \cdot \vec{a} }{c}  ,
\gamma^4 \frac{\vec{v} \cdot \vec{a} }{c^2}  \vec{v} + \gamma^2
\vec{a} \right)
\ee
Thus,
\begin{align} \label{a_a_prod}
a^{\alpha} a_{\alpha} &= -\gamma^6
\vec{a}^2\left[ 1 - \beta^2 \sin^2 \theta \right] \\ \nn
&= -\gamma^6 \left[\vec{a}^2 - \left| \frac{\vec{v} \times \vec{a}}{c} \right|^2 \right]
=
\begin{cases}
-\gamma^6 \vec{a}^2,& \vec{v} \parallel \vec{a} \\
-\gamma^4 \vec{a}^2,& \vec{v} \perp \vec{a}
\end{cases}\;.
\end{align}

\noindent
 This will change the non-relativistic Larmor power
formula given by Eq.\ (5), into the relativistic power, Li\'enard
formula \cite{lienard} which is,
\be \label{eq_lienard}
P_{\mbox{\tiny{Rel}}} = -\frac{ 2 e^2}{ 3 c^3 } a_\alpha a^\alpha
= -\frac{ 2 e^2}{ 3 c^3 } \frac{ d u_\alpha }{d\tau} \frac{ d
u^\alpha}{d \tau} = -\frac{2e^2}{3c^3}\gamma^6 \left( \vec{a}^2 -
\left| \frac{\vec v \times \vec a }{c} \right|^2 \right) =
-\frac{2e^2}{3c}\gamma^6 \left[{\dot{\beta}}^2 - \left(\vec \beta
\times \dot{\vec\beta}\right)^2\right] \;.
\ee
( We have used that $ d P^{\mu} / d\tau = a^{\mu}/m $, see Jackson \cite{jack}.)\\

\noindent
Now, we would like to use the same trick for the force term. We think it should generalize
in the same way.
First we show how to derive the full force term with the Poynting--Robertson ($dm/dt$) addition.
Take $\varepsilon$ as energy then we will equate energy loss $d\varepsilon/dt$ with force times velocity
in the usual way. Starting from,
\bea
\vec{F} &=& \frac{d\vec{p}}{dt} = \frac{d}{dt} (m \vec{v} ) = m \frac{d \vec{v}}{dt}
+ \frac{d m}{d t} \vec{v} \\ \nn
\frac{d \varepsilon}{dt} &=& \vec{F} \cdot \vec{v} = m \frac{d \vec{v}}{dt} \cdot \vec{v} + \frac{d m}{dt} \vec{v} \cdot \vec{v} \;\; .
\eea
We then time average our results assuming a periodic motion. We assume
that the velocity and acceleration will be the same at the start and finish of the
time integration, $v(t_1) = v(t_2)$ and $a(t_1) = a(t_2) $, then we equate the energy loss with the power radiated
given by the Larmor formula above.
\be
\int_{t_1}^{t_2} \left(m \frac{d\vec{v}}{dt} + \frac{d m}{dt} \vec v \right) \cdot \vec v dt
+ \frac{2 e^2}{3 c^3}\;\; \int_{t_1}^{t_2} \frac{d \vv }{dt} \cdot \frac{d \vv }{dt} dt = 0 \\ \nn
\ee
Integrate the power term by parts to find,
\be
\int_{t_1}^{t_2} \left[ m\frac{d\vec{v}}{dt} + R \frac{\vec v }{c^2} - \frac{2 e^2}{3 c^3}\dot{\vec a}
\right] \cdot \vec v dt = 0 \\ \nn
\ee
Now set the square bracket equal to zero, use $R$ as the Larmor power term,
and the new field (radiation) reaction force becomes,
\be
\vec{F}_{\mbox{\tiny Rad}} = -\frac{2 e^2}{3 c^3} \left[ \dot{\va} + \frac{\vv}{c^2} \vec{a}^2 \right]
\ee
This can now easily be generalized to the fully relativistic form by again setting
$\vec{a}^2 \rightarrow a^{\alpha} a_{\alpha}$, also $ d \va / dt \rightarrow d a_\mu /d\tau $ and
$ d \vv / dt \rightarrow d u_\mu / d\tau $ we find,
\be
f_\mu = \frac{2 e^2}{3 c^3} \left[ \frac{d a_\mu }{d\tau} + \frac{ u_\mu }{c^2} a^{\alpha}a_{\alpha} \right]
\ee
( We have used that $ d p^{\mu} /d\tau = a^{\mu}/m $ and
 $ d^2 P^{\mu} /d\tau^2 = \frac{1}{m}da^{\mu}/d\tau $,
see problems section in Jackson \cite{jack}.)\\

\noindent
We have used the Gaussian form here, since most of the older
papers we are citing are in those units. If you like to convert to S.I. units the constant term needs to be
multiplied by the factor $1/( 4 \pi \epsilon_0 )$.\\


\noindent
Below we generalize the above approach slightly to obtain the LAD equation directly from the
time average of the relativistic Larmor formula and Poynting--Robertson force. \\

\noindent
The equations of motion are
\be
f^\mu = \frac{dp^\mu }{d\tau}
\ee
with the spatial part
\bea
\vec{f} &=& \frac{ d \vec{p} }{d\tau} = m_0 \frac{ d( \gamma \vec{v})}{d\tau}
= m_0\gamma \frac{d \vec{v}}{d\tau} + \frac{ d(m_0  \gamma)}{d\tau} \vec{v} \nn \\
&=& m_0\gamma \frac{d \vec{v}}{d\tau} + \frac{1}{c^2} \frac{d \varepsilon}{d\tau} \vec{v}
= m_0\gamma^2 \frac{d \vec{v}}{dt} + \frac{1}{c^2} \frac{d \varepsilon}{dt} \vec{u}\;.
\eea
where $\varepsilon = \gamma m_0 c^2 $ is the relativistic energy of the particle. \\

\noindent
Equating the energy loss and radiated power and time averaging, as before, leads to
\bea
\int_{\tau_1}^{\tau_2}\left(\vec{f} - \frac{1}{c^2} \frac{d \varepsilon }{dt}\vec u - m_0\gamma^2 \frac{d \vec{v}}{dt}\right)\cdot \vec u d\tau &=& \\ \nn
\int_{\tau_1}^{\tau_2}\left(\vec{f}\cdot\vec u + \gamma P_{\mbox{\tiny Drag}} - \gamma P_{\mbox{\tiny Rad}}\right) d\tau &=& 0 \;.
\eea
The Poynting--Robertson drag \cite{rob} is given by Eq.\ (\ref{eq_poynting}) and the radiated power by the
relativistic Larmor-Li\'enard formula Eq.\ (\ref{eq_lienard}).
In the instantaneous comoving frame $d u^\mu/d\tau = \left( 0, d\vec{u}/d\tau\right)$ and the latter simplifies to
\be \label{eq_rad}
P_{\mbox{\tiny Rad}}= -\frac{2 e^2}{3 c^3 } \left( \frac{ d \vec{u}}{d\tau} \right)^2 \;.
\ee
Substituting from Eq.\ (\ref{eq_poynting}) and Eq.\ (\ref{eq_rad}),
\be
\int_{\tau_1}^{\tau_2}\left\{ \left[ \vec{f} + \frac{ 2 e^2}{ 3 c^5 }
\left(\frac{ d \vec{u}}{d\tau} \right)^2 \vec u \right] \cdot \vec u
+ \frac{2 e^2}{3 c^3 } \frac{ d \vec{u}}{d\tau} \cdot \frac{ d \vec{u}}{d\tau} \right\} d\tau = 0
\ee
and by partial integration of the second term,
\be \label{LAD_int}
\int_{\tau_1}^{\tau_2} \left( \vec{f} - \frac{2 e^2}{3 c^3 }\left[
\frac{d^2 \vec u }{d\tau^2} - \frac{1}{c^2}\left( \frac{ d \vec{u}}{d\tau} \right)^2 \vec u  \right] \right)
\cdot \vec u d\tau
+ \left[ \frac{2 e^2}{3 c^3 }\frac{ d \vec u }{d\tau}\cdot \vec u \right]_{\tau_1}^{\tau_2}
= 0 \;.
\ee
For periodic motion it follows that the LAD force \cite{dirac} is given by,
\be
f_i = \frac{ 2 e^2}{3 c^3 }\left [\frac{ d^2 \vec u }{d\tau^2} -
\frac{1}{c^2}\left( \frac{ d \vec{u}}{d\tau} \right)^2 \vec u \right] \rightarrow
\frac{2 e^2}{3 c^3} \left [ \frac{ d^2 u_\mu }{d \tau^2} +
\frac{1}{c^2} \frac{ d u^\alpha}{d \tau } \frac{ d u_\alpha}{d \tau} u_\mu  \right] \;.
\ee

\noindent
The first term is known as the Schott term which is an acceleration energy, \cite{schott, rohr97}.\\

\noindent
To emphasize the importance and  history of the Poynting--Robertson force term
we summarize several particle systems below, where this force term
naturally arises. Note the absence of the old $ \dot{a}$ force term.\\

\subsubsection*{The Rutherford--Bohr Hydrogen Atom Model Revisited}

It is interesting to note that a dust particle \cite{poy,rob} will spiral
into to sun due to the drag force acting upon it by the re--radiation of
the suns emissions. The exact same effect would be experienced by an electron
in a Keplerian orbit about a charge $Ze$. \cite{noon,larmor97,synge}.  This is described
by Synge \cite{synge} as the ``electromagnetic Kepler problem".
Following Synge, he starts out with the Lorentz equation of motion for an electron as
\be
m \frac{ d u_{\mu}}{d\tau} = \frac{e}{c}\tensor{F}{_\mu^\alpha} u_{\alpha}
\ee
where $u_{\mu} = dx_{\mu}/d\tau$ the proper velocity. Time dilation gives $d/dt = \g d/d\tau$.
The spatial components lead to a force equation while the
time component leads to a energy loss equation.
The force equation leads to an equation of motion similar to an ellipse,
 with precession of the apses. A special case of the ellipse is the
circle. The Lorentz equation of motion does not
show any sign of spiraling into the center as the charged particle radiates.
We must instead start with the LAD equation.
M. Abraham and M. von Laue  (see \cite{abraham,laue}) modified Lorentz's
equation of motion for an electron. This equation was obtained later by Dirac \cite{dirac}
via a fully covariant approach. The modified equation given by Synge is
\be \label{eq_Synge}
m \dot{u}_{\mu} = \frac{e}{c}\tensor{F}{_\mu^\alpha} u_{\alpha} + \frac{2}{3} \frac{e^2}{c^3}\left( \ddot{u}_{\mu}
- \dot{u}_{\alpha} \dot{u}^{\alpha}\frac{u_{\mu}}{c^2} \right)
= \frac{e}{c}\tensor{F}{_\mu^\alpha} u_{\alpha} + \frac{2}{3} \frac{e^2}{c^3} \dot{a}^\mu_\perp
\ee
where a dot denotes differentiation with respect to proper time $\tau$ and we have introduced
\be \label{a_dot_perp}
\dot{a}^\mu_\perp = \ddot{u}_{\mu} - \dot{u}_{\alpha} \dot{u}^{\alpha}\frac{u_{\mu}}{c^2}
\ee
defined by
\be
u_\alpha \dot{a}^\alpha_\perp = 0
\ee
since
\begin{subequations}
\begin{align}
u_\alpha u^\alpha &= -c^2 \;, \\
u_\alpha a^\alpha &= 0 \;, \\
u_\alpha \dot{a}^\alpha &= -a_\alpha a^\alpha \;.
\end{align}
\end{subequations}
In the local rest frame $u^\mu = \left( c, \vec{0}\right)$ and
$\dot{a}^\mu_\perp = \left( 0, \dot{\vec{a}}\right)$.
The last term in Eq. (20), can
be recognized as the Poynting--Robertson drag force. The temporal component
of this term, gives the Larmor power loss due to
radiation by the electron. Note that the mass $m$ here is the observed mass which is
the sum of the electromagnetic and bare masses.\\


\noindent
Rohrlich analyzed the LAD equation in curvilinear coordinates, i.e., with curvature $\kappa$ and torsion $\tau$. For example, a curve with constant curvature and torsion is a helix, which becomes a circle for zero torsion, and a straight line for zero curvature. For the Coulomb force problem in section 6-15 \cite{rohr} assuming circular motion, he finds the radiated power to be
\be \label{rad_Rohrich}
R = \frac{2c}{3} \frac{e^2}{r^2} (\gamma^2 -1)^2 = \frac{2c}{3} \frac{e^2}{r^2}\beta^4\gamma^4 = \frac{2}{3} \frac{e^2}{c^3} \gamma^4 \vec{a}^2
\ee
This is basically from the Larmor power since $(\gamma^2 -1) = \beta^2\gamma^2$ and for
circular motion $\left|\vec{a}\right| = v^2/r$. This power loss originates from the tangential component
\be
\frac{\vec F \cdot \vec v}{c} = \frac{2}{3}\frac{e^2}{r^2}\beta^2\gamma^2\left( \gamma^2-1\right) = \frac{2}{3}\frac{e^2}{r^2}\left( \gamma^2-1\right)^2
\ee
of the force experienced by a charge undergoing circular motion
\be
\vec{F} = -m\gamma \frac{v^2}{r} \hat{r} + \frac{2}{3}\frac{e^2}{r^2}\gamma^2\left( \gamma^2-1\right)\frac{\vec{v}}{c} \; .
\ee
Similarly, from Eqs. (\ref{eq_Synge}-\ref{a_dot_perp}) and (\ref{a_a_prod}) we obtain
\be
m \dot{u}_{\mu} = \frac{e}{c}\tensor{F}{_\mu^\alpha} u_{\alpha} + \frac{2}{3} \frac{e^2}{c^3} \dot{a}^\mu_\perp
= \frac{e}{c}\tensor{F}{_\mu^\alpha} u_{\alpha} + \frac{2}{3c} \frac{e^2}{r^2}\beta^4\gamma^4 u_{\mu}
\ee
since
\be
\dot{a}^\mu_\perp = \gamma^4 \vec{a}^2\frac{u_{\mu}}{c^2} =\frac{c^2}{r^2} \beta^4\gamma^4 u^\mu
\ee
and the temporal component (dividing through by $\g$ from the proper velocity terms) gives,
\be
\frac{\vec F \cdot \vec v}{c} = m \frac{du_0}{dt} = \frac{e}{\gamma c} \tensor{F}{_0^\alpha} u_{\alpha} + \frac{2}{3} \frac{e^2}{r^2}\beta^4\gamma^4
\ee
is the same as Eqs. (\ref{rad_Rohrich}) and (25). \\

\noindent
To summarize, an external tangential force is required to sustain circular motion for a point charge. However, this is not needed for a ring charge, i.e., with uniformly accelerated, periodic motion, and no contribution from the boundary terms. Simply put, a current loop with constant magnetic moment. \\

\noindent
For more of a history on the Kepler problem, Plass \cite{plass} gives a review of solutions to the Dirac equation.
At the end of his lengthy article he talks about the attractive (and repulsive) coulomb force.
His equation of motion for the attractive case, is his Eq.\ (162).
He takes the orbit to lie in the xy--plane and writes equations for x and y
time dependent motion, his Eq.\ (163). It is not possible to obtain exact
solutions to these equations although he numerically calculates curves and plots them in his Fig 10.
He references Clavier \cite{clavier}, who finds solutions in  one dimension and
also finds that in 3--D a logarithmic spiral satifies the equations at small
distances from the origin.

\subsubsection*{Einstein (and Hopf): Force on a 2--Level Atom}

Einstein and Hopf \cite{einhopf}, in (1910), derived the force acting on a dipole oscillator when
it is moving through an isotropic thermal field at velocity $v$. Later, more
clearly, Einstein 1917 derived a similar force acting on a 2--level atom when
it is moving with velocity $v$ in an isotropic thermal radiation field. For a modern account of
Einsteins 1917 work, see  Milonni \cite{mil} and a similar calculation was also
performed by Boyer \cite{boyer}.\\

\noindent
Einstein calculated the power radiated by
a 2--level atom by thermodynamic considerations and then using his force
equation solved for the equivalent power term and set the two terms equal.
Thus doing he solved the resulting first order differential eqn in
energy density $\rho(\w)$ and solved for the Planck distribution law.
According to Einstein \cite{einstein}, in the last statement of that paper,
\begin{quote}
``  a theory (of thermal radiation) can only be regarded as justified when it is
able to show that the impulses transmitted by the radiation field to
matter lead to motions that are in accordance with the theory of heat."
\end{quote}

\noindent
Einstein showed that the momentum transfer accompanying emission and
absorption of radiation are consistent with statistical mechanics if the
thermal radiation follows the Planck distribution. In so doing he derived a
force experienced by a particle moving through a thermal field,
\bea
F &=& -\frac{ \hbar \w'}{c^2} B_{12} \left( N_1 - N_2 \right)
\left( \rho (\w') - \frac{ \w'}{3} \frac{\p \rho }{\p \w' } \right) v \\
 &=& - \frac{1}{c^2} R v \\
\eea

\noindent
where $\w' $ is the Doppler shifted freq of the thermal radiation seen
by the moving atom, $B_{12}$ is the Einstein B--coefficient, $N_1$ and
$N_2$ are the populations of levels 1 and 2 inside the 2--level atom,
$v$ is the velocity of the atom and the power radiated $R$ becomes
\be
R= \hbar \w' B_{12} \left( N_1 - N_2 \right)
\left( \rho (\w') - \frac{ \w'}{3} \frac{\p \rho }{\p \w' } \right)
\ee
\noindent
We note that when we take the vacuum energy density
$ \rho (\w) = \hbar \w^3/(2 \pi^2 c^3)$ the term
\be
\left( \rho (\w') - \frac{ \w'}{3} \frac{\p \rho }{\p \w' } \right) = 0
\ee
\noindent
so there is no drag force on an atom in a vacuum. This appears to agree with
Page 1918 \cite{page}.\\

\subsubsection*{Dirac Equation of Motion of an Electron}

The main result of Dirac's \cite{dirac} paper is the equation of motion of
an electron, his Eq.\ (24). (You can find this referenced in the problems section of Jackson \cite{jack}.)
The force equation reads
\be
m \dot{u}_{\mu} - \frac{2}{3} \frac{e^2}{c^3}\left( \ddot{u}_{\mu}  -  \frac{1}{c^2} \dot{u}^2 u_{\mu}\right)
= \frac{e}{c} u_{\alpha}F_{\mu\;\mathrm{in}}^{\;\alpha}
\ee
The factors of $c$ have been written in, Dirac sets $c=1$ throughout his paper.
The 4--vector $u^{\mu}$ is $(c\dot{t}, \dot{x}, \dot{y}, \dot{z})$, and
$\dot{u}^2$ term is equivalent to
$ \dot{u}_{\alpha} \cdot \dot{u}^{\alpha} \equiv ( c^2 \ddot{t}^2 - \ddot{x}^2 )$
for motion along the x--direction only.
(Time derivatives are wrt. the proper time $\tau$ or $s$ in Dirac's paper.)
The first term clearly comes from the kinetic energy of the electron.
The second term on the
left looks like the Abraham Lorentz radiation reaction term
$F = 2e^2 \dot{a}/3 c^3$. The third term on the left is of the
form $F= -Rv/c^2$ where $R$ is the regular Larmor
formula for power radiated by an electron of acceleration $a$.
This clearly corresponds to the Robertson-Poynting force result.
 The right hand side of the Eq.\ (25), involves the electromagnetic field tensor in a mixed state.
$ F_{\mu\;\mathrm{in}}^{\;\nu} = ( F_{\mu \alpha} )_{\mathrm{in}} \; g^{\alpha \nu}$ where
for a flat space--time $g_{\alpha \beta} = g^{\alpha \beta}$ and the metric
takes the form $g_{00} = 1$, $g_{11} = -1$, $g_{22} = -1$, and $g_{33}=-1$.\\

\noindent
A history of the Dirac (or LAD) equation of motion is included in
Rohrlich \cite{rohr97,rohr0}. It can be found in Milonni's book \cite{mil} and
the book by Grandy \cite{grandy}. The remarkable thing about Grandy's book is that
on page 204, he actually writes the drag term in the LAD equation in the
Poynting--Robertson format $-Rv/c^2$ however no connection between this
radiation term and the Poynting--Robertson drag force is mentioned.

\subsubsection*{On the Unruh--Davies Effect}

Boyer \cite{boy80}, in (1980) noted that an electric dipole accelerated through the
vacuum would see a surrounding field not quite equal to the usual Planck
distribution. A correction term was needed, which turned out to be exactly
the radiation reaction term given by Poynting and Robertson, see \cite{boy84a}.
This showed that a
\begin{quote}
`` classical electric dipole oscillator accelerating though
classical electromagnetic zero--point radiation responds just as would a
dipole oscillator in an inertial
frame in a classical thermal radiation with Planck's
spectrum at temperature $T=\hbar a/2 \pi c k$"
\end{quote}

\noindent
where $T$ is the Unruh--Davies temperature.\\

\noindent
Boyer \cite{boy84b}, did a similar calculation for the spinning magnetic dipole
and found a mismatch with the Planck distribution again. He later corrected the magnetic
dipole work with a similar drag force to regain the Planck distribution,
[28]. This latter work refers to the classical theory of spinning particles
by Bhabha \cite{bhab}. Boyer \cite{boy84c} was able to show that,
\begin{quote}
 `` the departure from Planckian form
is cancelled by additional terms arising in the relativistic radiative damping
for the accelerating dipole. Thus the accelerating dipole behaves at
equilibrium as though in an inertial frame bathed by exactly Planck's spectrum
including zero--point radiation."
\end{quote}

\subsubsection*{Compton Scattering Force on an Electron}

In a very nice review article Blumenthal \cite{blum1} calculates synchrotron radiation, bremsstrahlung and
Compton forces of high energy electrons in dilute gases. He treats the Compton scattering in the ``rest frame"
of the electron, that is before electron has collided with a photon. From the electron rest frame it appears as though
the photons are approaching from a narrow cone. The direction of this cone would represent the direction of motion
of the electron in the lab frame. So a force due to the surrounding photons would appear, from the electron rest
frame, to be due to radiation pressure. Now from the lab frame, Blumenthal \cite{blum2} has a second shorter paper,
equally as good, treating the Poynting--Robertson force of the photons on the electrons in the lab frame.
From the lab frame the photons are incident from all directions making an angle $\theta$ with the direction of
motion of the electron. Blumenthal \cite{blum2}, calculated the mean force due to
Compton scattering on electrons with arbitrary velocity.  The electron was taken as moving along the
z--axis, the photons are in the xy--plane making angle $\theta$ with the z--axis. He works out the force in
all directions,
but we are specifically interested in the tangential force along the direction of motion of the electron.
The Poynting--Robertson tangential force, for light energy density $\rho$, is Eq.\ (19c) in
his paper, given as the rate of change of momentum $dP_3/dt$. We re-write this as,
\be
F_{\mbox{\tiny drag}} = - \gamma^2 \sigma_T \beta \int d \Omega \; \rho
F(\eta') (1 + \beta \cos \theta )^2 \frac{\sin \theta }{1 + \eta'}
\ee
here $\eta' = \g \rho ( 1 + \beta \cos \theta )/mc^2 $. For small $\eta'$ the function
$F(\eta') =1 - 16 \eta'/5 + \cdots $. For definitions of $F(\eta)$ see
the paper by Blumenthal \cite{blum2}.
We shall take $\eta'<< 1$, then $F(\eta') \approx 1$ and the equation above
can be greatly simplified.  Setting $\eta' \approx 0$ then
\bea
F_{\mbox{\tiny drag}} &=& - \gamma^2 \sigma_T \beta \int d \Omega \; \rho
(1 + \beta \cos \theta )^2 \sin \theta  \nn \\
 & \approx & - \sigma_T \frac{\beta}{c} \int d \Omega \; S (1 + 2 \beta \cos \theta ) \sin \theta
\eea
where $S= \rho c$ is the Poynting vector, to within a factor 1/4, $\beta = v/c$ and
$\sigma_T$ is the classical Thompson cross section (see Jackson \cite{jack}). We have also assumed
that $v << c$ so $\gamma \sim 1$ and terms in $v^2/c^2$ can be neglected. The Thompson
cross section is,
\be
\sigma_T = \frac{8 \pi}{3} \left(  \frac{ e^2}{m c^2} \right)^2 \;\; .
\ee
We note that $ e^2/ mc^2 = r_0$ is the classical electron radius.
\be
F_{\mbox{\tiny drag}} = - \left( \frac{ 8 \pi r_0^2}{3}\right)  S \frac{v}{c^2} \int_0^{2\pi} \int_0^{\pi/2} \;
(1 + 2 \beta \cos \theta ) \sin^2 \theta  d\theta \; d\phi \nn \\
\ee

\noindent
Integrating over the front half plane $ 0 < \theta < \pi/2$ we get,
\bea
F_{\mbox{\tiny drag}} &=& -\frac{4 \pi }{3} \left( 4 \pi r_0^2 \right) S
\left( \pi/4 + 2 \beta/3\right) \frac{v}{c^2} \nn \\
 &=& -\pi \frac{d \varepsilon}{dt} ( \pi/3 + 8 \beta/9 ) \frac{v}{c^2} \nn \\
 & \approx & -\pi \frac{d \varepsilon}{dt} ( 1 + \beta ) \frac{v}{c^2} \nn \\
 & \approx & -R \frac{v}{c^2} \; .
\eea
Here we have used  $d \varepsilon/dt$ as the energy loss and introduced the power loss $R =\pi  ( 1 + \beta ) d \varepsilon/dt$.
The electron radius has been used to define the spherical shell
$ 4 \pi r_0^2$ over which the Poynting vector is averaged.

\noindent
If instead you choose to integrate over $0 < \theta < \pi$ then the result is
\be
R = \frac{2 \pi^2}{3} 4 \pi r_0^2 S = \frac{ 2 \pi^2}{3} \frac{d \varepsilon}{dt}
\ee
Since we are evaluating the average Poynting vector $S$ at
a small radius $r_0$ from the electron, then not only do we pick up the
Compton scattered light intensity but also the Bremsstrahlung of the electron
no matter how small it is. So the rate of energy loss $d \varepsilon/dt$ above takes into
account both the scattered radiation and any electromagnetic radiation from
the electron because it is accelerating.

\subsubsection*{Conclusions}

We have shown that the addition of the Poynting--Robertson, energy loss term, into the standard
Abraham--Lorentz radiation reaction force term (in $\dot{a}$) seems appropriate. This is required
to account for numerous physical situations where a drag force is experienced and yet no $\dot{a}$ term is present.
Our main result of this paper is Eq.\ (10) and the preceding derivation and
the relativistic generalization following that leading to Eq.\ (18) which agrees with
the expression found in Jackson \cite{jack}, written in terms of the proper momentum $P^{\mu}$.
The Poynting--Robertson term has a long history. We have shown several instances where the term arises and
is needed to explain a physical effect. We believe by inclusion of the Poynting--Robertson term the
non--relativistic theory can be much more easily generalized to the relativistic form and takes on
the natural expected value for slower motion.\\
   Furthermore, one should note that the fully relativistic force
expression agreed upon by all recent text books was originally derived by Dirac (1938) \cite{dirac}
using both retarded and advanced waves.
It is difficult to find advanced waves mentioned in modern
text books, you would need to look in an
older electromagnetism book, for example, Panofsky and Phillips \cite{pan}.
Wheeler and Feynman elaborated on the advanced waves in 1945 \cite{wheeler} by introducing a physical mechanism,
the absorber, to account for the advanced waves.
We have derived the same terms without use of the advanced waves which appear
instead as a mass--energy change in the particle.\\

\end{document}